\documentclass[tradiabstract]{aa}

\newcommand{\bea}{\begin{eqnarray}}
\newcommand{\eea}{\end{eqnarray}}
\newcommand{\be}{\begin{equation}}
\newcommand{\ee}{\end{equation}}
\newcommand{\bc}{\begin{center}}
\newcommand{\ec}{\end{center}}
\newcommand{\ben}{\begin{enumerate}}
\newcommand{\een}{\end{enumerate}}
\newcommand{\bd}{\begin{description}}
\newcommand{\ed}{\end{description}}
\newcommand{\bmi}[1]{\begin{minipage}{#1 cm}}
\newcommand{\emi}{\end{minipage}}
\newcommand{\bmif}[1]{\begin{minipage}{#1\textwidth}}

\def\llabel#1{\label{sc:#1}  {#1}\hspace{0.5cm}}
\def\elabel#1{\label{eq:#1}\fbox{#1}}
\def\llabel#1{\label{sc:#1}}
\def\elabel#1{\label{eq:#1}}

\def\eck#1{\left\lbrack #1 \right\rbrack}

\def\rund#1{\left( #1 \right)}

\def\C{\tens{C}}
\def\D{{\cal D}}

\def\W{\tens{W}}

\def\d{{\rm d}}

\def\vp{\varphi}
\def\vt{{\vartheta}}

\def\Real{{\rm I\mathchoice{\kern-0.70mm}{\kern-0.70mm}{\kern-0.65mm}%
  {\kern-0.50mm}R}}
\def\bx#1{\leavevmode\thinspace\hbox{\vrule\vtop{\vbox{\hrule\kern1pt
        \hbox{\vphantom{\tt/}\thinspace{\bf#1}\thinspace}}
      \kern1pt\hrule}\vrule}\thinspace}

\def\vc#1{{\mbox{\boldmath$#1$\unboldmath}}}
{\catcode`\@=11
\gdef\SchlangeUnter#1#2{\lower2pt\vbox{\baselineskip 0pt \lineskip0pt
  \ialign{$\m@th#1\hfil##\hfil$\crcr#2\crcr\sim\crcr}}}
}

\def\ueber#1#2{{\setbox0=\hbox{$#1$}%
  \setbox1=\hbox to\wd0{\hss$\scriptscriptstyle #2$\hss}%
  \offinterlineskip
  \vbox{\box1\kern0.4mm\box0}}{}}

\def\bx#1{\leavevmode\thinspace\hbox{\vrule\vtop{\vbox{\hrule\kern1pt
        \hbox{\vphantom{\tt/}\thinspace{\bf#1}\thinspace}}
      \kern1pt\hrule}\vrule}\thinspace}

{\catcode`\@=11
\gdef\SchlangeUnter#1#2{\lower2pt\vbox{\baselineskip 0pt \lineskip0pt
  \ialign{$\m@th#1\hfil##\hfil$\crcr#2\crcr\sim\crcr}}}
}

\def\ts{\thinspace}

\usepackage{graphicx}
\usepackage{txfonts}
\usepackage{natbib}
\usepackage{subfigure}
\usepackage{enumerate}
\usepackage{slashbox}
\begin{document}

   \title{Generalized multi-plane gravitational lensing: time delays,
     recursive lens equation, and the mass-sheet transformation}

   \author{Peter Schneider \inst{1}  %\& Dominique Sluse
          %\inst{1}
          }

   \institute{Argelander-Institut f\"ur Astronomie, Universit\"at
     Bonn, Auf dem H\"ugel 71, D-53121 Bonn, Germany\\
    peter@astro.uni-bonn.de}

%   \date{Received September 15, 1996; accepted March 16, 1997}

% \abstract{}{}{}{}{} 
% 5 {} token are mandatory
 
  \abstract
  {We consider several aspects of the generalized multi-plane
    gravitational lens theory, in which light rays from a distant
    source are affected by several main deflectors, and in addition by
    the tidal gravitational field of the large-scale matter
    distribution in the Universe when propagating between the main
    deflectors. Specifically, we derive a simple expression for the
    time-delay function in this case, making use of the general
    formalism for treating light propagation in inhomogeneous
    spacetimes which leads to the characterization of distance
    matrices between main lens planes. Applying Fermat's principle, an
    alternative form of the corresponding lens equation is derived,
    which connects the impact vectors in three consecutive main lens
    planes, and we show that this form of the lens equation is
    equivalent to the more standard one. For this, some general
    relations for cosmological distance matrices are derived. The
    generalized multi-plane lens situation admits a generalized
    mass-sheet transformation, which corresponds to uniform isotropic
    scaling in each lens plane, a corresponding scaling of the
    deflection angle, and the addition of a tidal matrix (mass sheet
    plus external shear) to each main lens. We show that the time
    delay for sources in all lens planes scale with the same factor
    under this generalized mass-sheet transformation, thus precluding
    the use of time-delay ratios to break the mass-sheet
    transformation.  }

   \keywords{cosmological parameters -- gravitational lensing: strong 
               }
  \titlerunning{Multi-plane gravitational lensing}

   \maketitle

\section{\llabel{Sc1}Introduction}
In strong gravitational lensing systems, in which a galaxy or a galaxy
cluster causes multiple images or strong image distortions of
background sources, one often neglects the inhomogeneities of the
gravitational field between the observer and the lens, and between the
lens and the source \citep[see, e.g.,][for reviews on strong lensing
systems]{SaasFee3,Treu2010,Bart10,KneibNata11}. This usually provides
a very good approximation, since the lensing strength of the main lens
over the region where strong lensing effects occur is much larger than
the typical distortion effects of matter along the line-of-sight. The
latter is comparable to the typical strength of cosmic shear effects
\citep[see, e.g.,][]{bs-review,sch06-WL,Munshi08,Hoek13}, and amounts to about
1 or 2 percent of the distortion in the strong-lens region.

Whereas the propagation effects are thus small, the interest of these
weak distortions has been renewed, for at least two different
reasons. The first is that strong lens systems may be biased towards
showing relatively strong line-of-sight structures, thereby increasing
their lensing probability. This effect is most likely affecting the
lensing cross sections for the formation of giant arcs in clusters
(see Bartelmann et al.\ \citeyear{Bart98} for stating the `arc
statistic problem', and Meneghetti et al.\ \citeyear{Meneg13} for a recent
update of the issue). Cosmological simulations \citep[][and references
therein]{PuchHilb09} indicate that line-of-sight structure can indeed
substantially modify the lensing efficiency of clusters. More
recently, \cite{Bayliss14} found a significant overdensity of galaxy
groups along the line-of-sight towards strong-lensing clusters,
observationally supporting the presence of this bias.

The second reason for the renewed interest in intervening distortions
of strong-lens systems is their use for a precise determination of
parameters, most noticeable the Hubble constant from measured time
delays \citep{Refsdal1964} in multiply imaged QSOs \citep[see,
e.g.,][]{Koch03,SaasFee3,Treu2010,Suyu2010,Courbin11,Suyu2013a,Tewes13}. The
quality of 
modern imaging data and the accuracy of time delay estimates allows
one to derive estimates of the Hubble constant with a formal error of
$\sim 6\%$ \citep{Suyu2013a}. At this level of precision, line-of-sight
effects may become highly relevant in these strong-lensing systems
\citep[e.g.,][and references therein]{Wong11,Coll13,Greene13}.

With additional deflectors along the line-of-sight, the lens equation,
which relates the true source position to the observed positions of
images, needs to be generalized to include deflection at more than one
distance from the observer. \cite{BlandNara86} % and \cite{SchnBorg86}
established the theory of multi-plane gravitational lensing \citep[see also
Chap.\ts 9 of][]{SEF}, where the mapping between images and their
source is affected by the action of several deflectors at different
redshifts. This multi-plane lensing is employed in ray-tracing
simulations -- see \cite{Refs70} for the earliest ray-tracing
simulations, \cite{JSW00}, \cite{HHWS09}, and references therein --
where the three-dimensional mass distribution between observer and
source is partitioned into separate slices, and the mass distribution
in each slice is treated as a gravitational lens plane. 

The full theory of multiple deflection lensing is required if there
are two or more strong deflectors along the same line-of-sight towards
a background source. For galaxy-scale lensing, such systems are rare,
since such an alignment is not very probable. However, in the sample
of several hundred galaxy-scale strong lens systems currently known, there are
examples of multiple lenses at two different redshifts
\citep{Chae01,Gavaz08}. These systems may be of particular interest,
since they may be used in principle to determine cosmological distance
ratios \citep{CA14}, and therefore to constrain the cosmic expansion
history, although in practice the mass-sheet transformation renders
this a difficult task \citep{S14}. 

Far more common are situations where there is a single strong lens
in the line-of-sight to a distant source, and several other deflectors
at different redshifts situated sufficiently far away from the
strong-lensing region of the main deflector such that their deflection
angle can be linearized across this strong-lensing
region. \cite{Kovner87} considered this case of one main lens,
combined with linearized deflections at different redshifts. The
effects of the linear deflectors can be summarized into a set of
matrices which describe the mapping between angles at the vertices of
light cones to the separation vectors along the light cone at the
redshifts of lens and source. \cite{Schn97} reconsidered this
situation, using the general formalism of light propagation in an
inhomogeneous universe \citep[][hereafter SSE]{SSE}, and related this
generalized 
gravitational lens equation to the one where the deflection occurs
in only a single plane, with a tidal deflection matrix added
\citep[see also][]{1997ApJ...482..604K}. 
\cite{McCully14} further generalized this theory to consider several
main lenses, together with linearized deflections between the main
lens planes. In particular, \cite{McCully14} emphasized the advantages
of this hybrid framework for the modeling of strong lens systems with
multiple main deflectors along the line-of-sight.

One of the prime motivations for the work of \cite{McCully14} was the
derivation of the time-delay function for such generalized lensing
systems, as this is required for relating the time delay to the
scale-length of the Universe, i.e., the Hubble radius. Using the
Millennium Simulation, \cite{2014MNRAS.439.2432J} investigated the
impact of the line-of-sight matter distribution on strong lensing
properties, including the time delay; they concluded that the
intermediate mass distribution leads to a spread of $\sim 6\%$ in the
product of Hubble constant and time delay, for strong lensing systems
with source redshifts $z_{\rm s}\sim 2$.  Arguably, the largest
obstacle for the time-delay method to obtain accurate estimates for
the Hubble constant is the degeneracy of the mass model due to the
mass-sheet transformation (MST; Falco et al.\ \citeyear{FGS85}; see
Schneider \& Sluse \citeyear{SS13} for a recent discussion) or the
more general source position transformation \citep{SS14}. Recently,
\cite{S14} has shown that an MST also exists for the case of lenses at
two different distances from us. In particular, this MST leads to a
scaling of all time delays (from sources on both source planes) by the
same factor, thus precluding that the degeneracy due to the MST can be
broken by time-delay ratios.

In this paper, these results will be further generalized to the case
of arbitrarily many main lens planes, together with linear deflections
between the main lens (and source) planes. After a short summary of
the propagation equations in an inhomogeneous universe, as they apply
to the case under consideration here, we derive the time-delay
function for the case of several main lens planes. Whereas this has
been done before by \cite{McCully14}, our expression for the light
travel time function is expressed in a form which allows us to apply
Fermat's theorem in gravitational lensing. Thus, the lens equation can
be obtained from requiring that the light travel time function is
stationary with respect to all impact vectors in the main lens
planes. With this procedure, we derive an alternative form of the lens
equation, which involves the impact vectors of three consecutive lens
planes and the deflection angle in the middle one of them. We show
that this iterative form of the lens equation is equivalent to the
more standard form; in order to do so, we first obtain a very general
relation between distance matrices.

We then turn to the MST in this case, and find a curious property of
its behavior: The uniform, isotropic scaling factor which
characterizes the MST alternates between a free parameter $\lambda$
and unity from one lens/source plane to the next. The corresponding
modification of the deflection angle in the main lens planes
corresponds to a scaling of the deflection, plus the addition of a
tidal deflection matrix (which in the absence of linear deflections
between lens planes reduces to the addition of a uniform mass sheet). 
Finally we show that all time delays, for sources located on any plane,
scale by the same factor under an MST, precluding the possibility of
breaking the degeneracy from the MST by measured time-delay ratios.

\section{Generalized multi-plane lens equation}
\subsection{Optical tidal equation}
The propagation of light rays follows from the geodesic equation,
specialized to a perturbed Robertson--Walker metric \citep[see][and
references therein]{SEF,SSE,Bart10}. In particular, for infinitesimally
small light bundles, the separation vector $\vc\xi$ of a ray from the
reference (or `central') ray of the bundle is given by the optical
tidal equation
\be
{\d^2 \vc\xi(\lambda)\over \d\lambda^2}=\cal{T}(\lambda) \,\vc\xi(\lambda)
\elabel{I1}
\ee
where $\cal{T}$ is the optical tidal matrix, evaluated at the affine
parameter $\lambda$ along the central ray. The affine
parameter $\lambda$ is
related to the redshift $z$, the cosmic scale factor $a=1/(1+z)$, or
cosmic time $t$ through
\be
\d\lambda = -{c\,\d t \over (1+z)}=-{c\,\d a\over H}
={c\,\d z\over H\,(1+z)^2} \;,
\elabel{I2}
\ee
where we have assumed that $\lambda$ increases with redshift, i.e.,
decreases with cosmic time, 
\be
H=\dot a/a= H_0\sqrt{\Omega_{\rm
    m}(1+z)^3+(1-\Omega_{\rm m}-\Omega_\Lambda)(1+z)^2+\Omega_\Lambda}
\elabel{I2a}
\ee
is the Hubble function,
$H_0$ is the Hubble constant, and $\Omega_{\rm m}$ and
$\Omega_\Lambda$ are the cosmic density parameters in matter and
vacuum energy.

We consider $\cal{T}$ to consist of three separate components, ${\cal
  T}= {\cal T}_{\rm bg}+{\cal T}_{\rm sm}+{\cal T}_{\rm cl}$. The first is
the optical tidal matrix of the homogeneous background universe, and
is given by (see SSE)
\be
{\cal T}_{\rm bg}=-{3\over 2}\rund{H_0\over c}^2 \Omega_{\rm
  m}(1+z)^5\cal{I}\;,
\elabel{I3}
\ee
where $\cal{I}$ is the two-dimensional unit matrix. If we consider a
light bundle with vertex at redshift $z=0$ and affine parameter $\lambda=0$,
which is only subject to ${\cal T}_{\rm bg}$, then the solution of
(\ref{eq:I1}) is $\vc\xi(\lambda)=D(\lambda)\vc\theta$, where
$\vc\theta$ is the angle that the light ray under consideration
encloses with the fiducial ray at the vertex, and $D(\lambda)$ is the
solution of the differential equation
\be
{\d^2 D(\lambda)\over \d\lambda^2}=-{3\over 2}\rund{H_0\over c}^2 \Omega_{\rm
  m}(1+z)^5\,D(\lambda)\;,
\elabel{I4}
\ee
with initial condition $D(0)=0$ and $\d D/\d\lambda =1$. $D$ is the
angular-diameter distance of the homogeneous universe, as a function
of affine parameter (or redshift). 

We consider two different kinds of inhomogeneities here. The first of
them is related to small-scale density inhomogeneities, such as
galaxies and their dark matter halos. For those, the optical tidal
matrix ${\cal T}_{\rm cl}$ is a strong function of position, and there
is no longer a linear relation between enclosed angle $\vc\theta$ and
separation $\vc\xi$ of a light ray, for finite $\vc\theta$. In
strong-lensing applications, we are typically interested only in that
region of these small-scale inhomogeneities where multiple images can
be formed, which corresponds to a few tens of kiloparsecs for
galaxy lenses. These
small-scale inhomogeneities will be considered explicitly as `main
lenses' in the following. The second kind of inhomogeneities is due to
the large-scale mass distribution of the universe. We assume that the
corresponding gravitational field is sufficiently smooth, so that the
tidal effects can be considered approximately constant across the
region where strong-lensing effects occur. Hence we assume that over
such a region, ${\cal T}_{\rm sm}$ can be considered to depend only of
$\lambda$, not on the actual position of the light ray. According to
SSE, this contribution of the optical tidal matrix is 
\be
\rund{{\cal T}_{\rm sm}}_{ij}=-{(1+z)^2\over c^2}
\rund{2{\partial^2\phi\over\partial \xi_i\,\partial\xi_j}
+\delta_{ij}{\partial^2\phi\over\partial \xi_3^2}}\;,
\elabel{I4a}
\ee
where $\phi$ is the Newtonian potential sourced by the density
inhomogeneity, i.e., satisfying the Poisson equation $\nabla_\xi^2
\phi=4\pi G (\rho-\bar\rho)$, where 
$\bar\rho(z)$ is the mean matter density in the Universe, and we
assumed that the light ray propagates in the $\xi_3$-direction.

\subsection{Generalized multi-plane lens equation}
We consider a direction in the sky which has a set of main lenses along
the line-of-sight to a distant source, located at redshifts $z_i$ or
affine parameters $\lambda_i$. As shown in SSE \citep[see
also][]{Schn97}, the separation vector $\vc\xi(\lambda)$ then becomes 
\be
\vc\xi(\lambda)=\D (\lambda)\,\vc\theta - 
\sum_i
\D_i(\lambda)\eck{\hat{\vc\alpha_i}(\vc\xi_i)-\hat{\vc\alpha}_i^0}
{\rm  H}(\lambda-\lambda_i)\;,
\elabel{I5}
\ee
where $\vc\theta$ is the angle the light encloses with the fiducial
ray at the observer. The distance matrix $\D(\lambda)$ solves the
optical tidal equation
\be {\d^2\D(\lambda)\over \d\lambda^2}=\eck{{\cal T}_{\rm
    bg}(\lambda)+{\cal T}_{\rm sm}(\lambda)} \D(\lambda) \elabel{I6}
\ee
with initial conditions $\D(0)=0$ and $(\d\D/\d\lambda)(0)={\cal I}$, and the
distance matrices $\D_i(\lambda)$ solve the same differential
equation, but with initial conditions $\D_i(\lambda_i)=0$ and
$(\d\D_i/\d\lambda)(\lambda_i)=(1+z_i){\cal I}$. They are the distance
matrices which 
apply for light rays having their vertex at $\lambda_i$. In
(\ref{eq:I5}), ${\rm H}(\lambda-\lambda_i)$ is the Heaviside step
function. The deflection angle $\hat{\vc\alpha}_i(\vc\xi_i)$ is given
in terms of the surface mass density $\Sigma_i(\vc\xi_i)$ as
\be
\hat{\vc\alpha}_i(\vc\xi_i)={4G\over c^2}\int\d^2\xi'\;
\Sigma_i(\vc\xi'){\vc\xi_i-\vc\xi'\over |\vc\xi_i-\vc\xi'|^2}\;,
\elabel{I7}
\ee
and $\hat{\vc\alpha}_i^0$ denotes the deflection angle of the fiducial ray
in the $i-$th lens plane. If we now perform the translation
$\vc\xi(\lambda)= \tilde{\vc\xi}(\lambda) +\vc\eta(\lambda)$, with
\[
\vc\eta(\lambda)= \sum_i
\D_i(\lambda)\,\hat{\vc\alpha}_i^0\,
{\rm  H}(\lambda-\lambda_i)\;,
\]
then $\tilde{\vc\xi}$ satisfies (\ref{eq:I5}) with the $\hat{\vc\alpha}_i^0$ set
to zero. In the following, we will always assume this (unobservable)
translation, and drop the tilde on $\tilde{\vc\xi}$ henceforth. For the
impact vectors $\vc\xi_j$ in the $j$-th plane, we then obtain
\be
\vc\xi_j=\D(\lambda_j)\,\vc\theta-
\sum_{i=1}^{j-1}\D_i(\lambda_j)\,\hat{\vc\alpha}_i(\vc\xi_i)
\equiv \D_j\, \vc\theta-
\sum_{i=1}^{j-1} \D_{ij}\,\hat{\vc\alpha}_i(\vc\xi_i)
\; .
\elabel{LEorig}
\ee

\subsection{Calculation of the distance matrices}
The distance matrices $\D_i(\lambda)$ depend on the large-scale matter
distribution around the line-of-sight, which is dominated by dark
matter and thus difficult to determine observationally. One
possibility to estimate ${\cal T}_{\rm sm}$ from observations is to
assume that galaxies provide a good tracer of the total matter
distribution on large scales. From an observed distribution of
galaxies (or a particular kind of galaxies, like luminous red
galaxies) around the line-of-sight, an estimate of the tidal field can
be obtained; this is the strategy proposed in \cite{Coll13} and
\cite{SBN14}. 

The propagation of light through the large-scale structure is the
subject of cosmological weak lensing, or cosmic shear \citep[see,
e.g.,][]{sch06-WL}. In cosmic shear, one usually describes the
propagation matrices in comoving coordinates. In order to connect the
formulation given here, where the $\D_i$ relate angles to proper
transverse separation vectors (which is appropriate, as the
matter distribution of the main lenses -- galaxies or clusters -- are
most conveniently described in physical scales), to that used in weak
lensing, we show in the appendix that 
\be
\D_i(\chi)=D_i(\chi){\cal I}
-{2\over c^2}\int_{\chi_i}^\chi \d\chi'\;
{a(\chi)\over a(\chi')} f_k(\chi-\chi')\tens{H}(\phi(\chi'))
\D_i(\chi')\;,
\elabel{I8}
\ee
where $\chi$ is the comoving distance, $f_k(\chi)$ is the comoving
angular diameter distance, $D_i(\chi)=a(\chi)f_k(\chi-\chi_i)$,
which satisfies the differential equation
\be
{\d^2 f_k(\chi)\over \d \chi^2}=-K\,f_k(\chi)
\elabel{I9}
\ee
with initial conditions $f_k(0)=0$ and  $\d f_k(0)/\d\chi=1$.
Furthermore, 
$K=(\Omega_{\rm m}+\Omega_\Lambda-1)H_0^2/c^2$ is the spatial
curvature of the universe, and $\tens{H}(\phi)$ is the two-dimensional
Hessian of the
gravitational potential $\phi$, evaluated in comoving transverse
coordinates. Provided the perturbations are small, so that $\D_i$
deviates only slightly from $D_i{\cal I}$, we can replace
$\D_i(\chi')$ by $D_i(\chi')$ in the integrand,
\be
\D_i(\chi)=D_i(\chi){\cal I}
-{2\over c^2}\int_{\chi_i}^\chi \d\chi'\;
{a(\chi)\over a(\chi')} f_k(\chi-\chi')\tens{H}(\phi(\chi'))
D_i(\chi')\;.
\elabel{I9a}
\ee
In weak lensing, this
approximation is called `the neglect of
lens-lens coupling', often also termed as `Born-approximation'.
This approximation is very accurate and certainly sufficient for the
purposes discussed in the current context.\footnote{It must be
  stressed here that this Born-approximation only applies to the
  distance matrices $\D_i$ which are governed by the smooth part of
  ${\cal T}$ according to (\ref{eq:I6}); only this smooth contribution
  is contained in (\ref{eq:I8}). No such approximation is made with
  regards to the main deflectors.}
Thus, the deviation of
$\D_i$ from $D_i{\cal I}$ is given as a line-of-sight integral over
the tidal force field, with a distance-dependent weighting. We also
note that in the approximation (\ref{eq:I9a}), the distance matrices
$\D_i$ are symmetric, which is generally not the case if the exact
expression (\ref{eq:I8}) is used.

For statistical studies, instead of trying to obtain the tidal field
along the line-of-sight towards the sources from observations of the
galaxy distribution in those directions, one can also derive the
probability distribution for the distance matrices from cosmological
simulations, as has been done in \cite{Suyu2013a}.

\section{Time-delay function and the iterative lens equation}
In this section, we first derive the light travel time function
(LTTF) corresponding to the generalized multi-plane lens equation
considered in the previous section. An alternative form of the lens
equation is then derived from Fermat's principle, which in the current
context states that
the lens equation is equivalent to setting the gradient of the LTTF
with respect to all impact vectors equal to zero (i.e., real light
rays correspond to stationary points of the LTTF). Then we show that
this iterative form of the lens equation is equivalent to
(\ref{eq:LEorig}). 

\subsection{Time-delay function}
We first consider a single lens plane at $z_1$ with a source at $z_2$,
in which case the lens equation reads
\be
\vc\xi_2=\D_2\,\vc\theta-\D_{12}\, \hat{\vc\alpha}(\vc\xi_1)
=\D_2\,\D_1^{-1}\vc\xi_1-\D_{12}\, \hat{\vc\alpha}(\vc\xi_1)\;,
\elabel{LE-1pl}
\ee
where we set
$\hat{\vc\alpha}(\vc\xi_1)\equiv\hat{\vc\alpha}_1(\vc\xi_1)$. 
We define the LTTF $\tau(\vc\xi_1,\vc\xi_2)$ to be the
excess light travel time from a source at $\vc\xi_2$ to the observer
caused by the deflection in the lens
plane at $\vc\xi_1$. This excess travel time has two components, a
geometrical one 
(a bent ray is longer than an unbent one), and a potential one caused
by the retardation of photons in the gravitational potential of the
deflector. 

In standard lens theory, with unperturbed angular-diameter distances $D_{ij}$,
the potential part of the time delay function takes the form 
$c\tau_{\rm pot}=-(1+z_1)(D_1 D_2/D_{12})\psi(\vc\theta)$, where
$\psi(\vc\theta)$ is 
the deflection potential, which satisfies
$\nabla_{\vc\theta}\psi=\vc\alpha(\vc\theta)\equiv
(D_{12}/D_2)\hat{\vc\alpha}(D_1\vc\theta)$. Now we define the potential
$\hat\psi(\vc\xi_1)$ such that it satisfies
$\nabla_{\vc\xi_1}\hat\psi=\hat{\vc\alpha}$. This potential is a multiple
of $\psi$, i.e., $\hat\psi(\vc\xi_1)=k\,\psi(\vc\xi_1/D_1)$, up to an
irrelevant additive constant. To find $k$,
we consider
\[
\hat{\vc\alpha}=\nabla_{\vc\xi_1}\hat\psi={k\over
  D_1}\nabla_{\vc\theta}\psi
={k\over D_1}\vc\alpha = {k\,D_{12}\over D_1 D_2}\hat{\vc\alpha}\;,
\]
yielding
\be
\hat\psi(\vc\xi_1)={D_1 D_2\over D_{12}}\psi(\vc\xi_1/D_1)\;.
\elabel{psipsi}
\ee
Hence, the potential part of the LTTF takes the form
$c\tau_{\rm pot}=-(1+z_1)\hat\psi(\vc\xi_1)$. As expected, it does not
depend on the cosmological distances, since it is caused solely by the
local effect of propagating through a gravitational field, and the
corresponding time interval is then redshifted by a factor $(1+z_1)$
to the observer. Accordingly, also in the case of perturbed light
propagation between the lens planes, the potential part of the time
delay must have the same form, since it is unaffected by propagation
effects. 

We find an explicit expression for $\tau(\vc\xi_1,\vc\xi_2)$ by
requiring that $\nabla_{\vc\xi_1} \tau(\vc\xi_1,\vc\xi_2) =0$ is
equivalent to the lens equation (\ref{eq:LE-1pl}). This fixes $\tau$
up to a multiplicative constant and terms which depend solely on
$\vc\xi_2$. The multiplicative constant is fixed by the explicit
expression given above for the potential part of $\tau$, and the
additive constant (which is irrelevant for time delay measurements) is
fixed by requiring that the geometrical part of $\tau$ should vanish
if the light ray is undeflected by the main lens. This then yields
\bea
c \tau(\vc\xi_1,\vc\xi_2) \!\!\!& = &\!\!\! {1+z_1 \over 2} 
\rund{\C_{12}\vc\xi_1 - \D_{12}^{-1}\vc\xi_2}^{\rm t}
\C_{12}^{-1}\rund{\C_{12}\vc\xi_1 - \D_{12}^{-1}\vc\xi_2}
\nonumber \\
&-&\!\!\!  (1+z_1)\,\hat\psi(\vc\xi_1) \;,
\elabel{tau1}
\eea
where we defined
\be
\C_{ij}=\D_{ij}^{-1}\D_j\D_i^{-1}\;.
\elabel{Cdef}
\ee
As shown in \cite{Schn97} -- see also \cite{Kovner87} -- the matrix $\C_{ij}$
is symmetric.
Indeed, we see that
\be
\nabla_{\vc\xi_1}\tau(\vc\xi_1,\vc\xi_2)={1+z_1\over c}
\rund{\C_{12}\vc\xi_1 - \D_{12}^{-1}\vc\xi_2-\hat{\vc\alpha}} =0
\elabel{tau-1p}
\ee
is equivalent to the lens equation (\ref{eq:LE-1pl}), as is easily
verified by multiplying the foregoing expression by $\D_{12}$ from the
left. We also note that $\tau=0$ if $\hat{\psi}=0$ and if 
the ray is unbent, i.e., if $\vc\xi_2=\D_2 \D_1^{-1}\vc\xi_1$, as we
required for $\tau$. \cite{McCully14} obtained a somewhat different form for
$\tau$, which they showed to be equivalent to the expression given
here. However, it must be pointed out that this equivalence applies
only to physical light rays, i.e., those which satisfy the
gravitational lensing equation (\ref{eq:LE-1pl}). For those rays, we
could write the light travel time as
\[
c\tau={1+z_1\over 2}\hat{\vc\alpha}_1^{\rm
  t}\tens{C}_{12}^{-1}\hat{\vc\alpha}_1-(1+z_1)\hat\psi(\vc\xi)\;,
\]
where the lens equation (\ref{eq:LE-1pl}) was used to eliminate
$\vc\xi_2$. However, (\ref{eq:tau1}) is more general, as it yields the
light travel time for all kinematically possible rays, not only for
those for which the bend by the main lenses equals the actual
deflection angle as calculated as the gradient of $\hat\psi$. This
more general form of the $\tau$ is needed if the lens equation is to
be derived from Fermat's principle.

In case of several main lens planes, the LTTF is obtained by
considering the replacement of the actual light ray by successively
straighter rays, i.e., by removing the bends of the ray and the
gravitational potentials they traverse \citep[see Sect.\ts 9.2
of][]{SEF}.  Removing the bend and deflection potential in the first
plane leads to the contribution (\ref{eq:tau1}) of the
LTTF. Subsequent removal of the bend and potential in the second lens
plane yields a similar contribution, with the indices (1,2) replaced
by (2,3). Iterating this consideration, we obtain for the general case
\bea
c \!\!\!\!\!\!&&\!\!\!\!\!\!\!\!\!\!\tau(\vc\xi_1,\vc\xi_2,\dots,\vc\xi_{N+1}) =
\sum_{i=1}^N (1+z_i) 
\elabel{tau}
\\
\!\!\!\!\!\!\!\!\!& \times &\!\!\!\!\!\! \Biggl[
{1\over 2} 
\rund{\C_{i,i+1}\vc\xi_i - \D_{i,i+1}^{-1}\vc\xi_{i+1}}^{\rm t}
\C_{i,i+1}^{-1}\rund{\C_{i,i+1}\vc\xi_i - \D_{i,i+1}^{-1}\vc\xi_{i+1}}
-  \hat\psi_i(\vc\xi_i) \Biggr] \nonumber 
\eea
as the sum over terms of the form (\ref{eq:tau1}) for the individual
planes. 

Any ray connecting the source at $\vc\xi_{N+1}$ and the observer is
fully characterized by the impact vectors $\vc\xi_i$, $1\le i\le N$ in
the lens planes, since between the planes, it follows the propagation
equation (\ref{eq:I1}) whose solution is uniquely determined by the
two impact vectors at consecutive planes. Therefore, the actual light
rays are singled out as those for which the LTTF is stationary, with
respect to variations of the impact vectors in the $N$ lens planes.
Thus,
the lens equation is obtained by setting the derivative of $\tau$
with respect to the $\vc\xi_j$ equal to zero. For each $j\ge 2$, two
terms of the above sum will contribute, namely the terms $i=j$ and
$i=j-1$. We obtain
\bea
\nabla_{\vc\xi_j}(c\tau)\!\!\!&=&\!\!\!(1+z_j)\eck{\C_{j,j+1}\vc\xi_j -
  \D_{j,j+1}^{-1}\vc\xi_{j+1}-\hat{\vc\alpha}_j(\vc\xi_j)} \nonumber \\
\!\!\!&+&\!\!\! (1+z_{j-1})\rund{\D_{j-1,j}^{-1}}^{\rm t}
\rund{\C_{j-1,j}^{-1}\D_{j-1,j}^{-1}\vc\xi_j-\vc\xi_{j-1}}=0\;,
\eea
or
\bea
\vc\xi_{j+1}\!\!\!&=&\!\!\!\D_{j,j+1}\eck{\C_{j,j+1}
+{1+z_{j-1}\over 1+z_j}\rund{\D_{j-1,j}^{-1}}^{\rm
  t}\C_{j-1,j}^{-1}\D_{j-1,j}^{-1}}\vc\xi_j \nonumber \\
\!\!\!&-&\!\!\!
{1+z_{j-1}\over 1+z_j}\D_{j,j+1}\rund{\D_{j-1,j}^{-1}}^{\rm
  t}\vc\xi_{j-1}-\D_{j,j+1}\hat{\vc\alpha}_j(\vc\xi_j) \;.
\elabel{LEnew1}
\eea
This equation relates the position vectors in three consecutive planes
to the deflection angle in the middle plane, quite in contrast to the
lens equation (\ref{eq:LEorig}) which contains all impact vectors
$\vc\xi_i$ for a given $\vc\xi_j$, $1\le i\le j-1$. Hence, this new
lens equation is more `local' than the original one. 

In the following we will explicitly show that these two forms of the
lens equation are equivalent. For this, we first need to derive a
general relation between distance matrices.

\subsection{A relation between distance matrices}
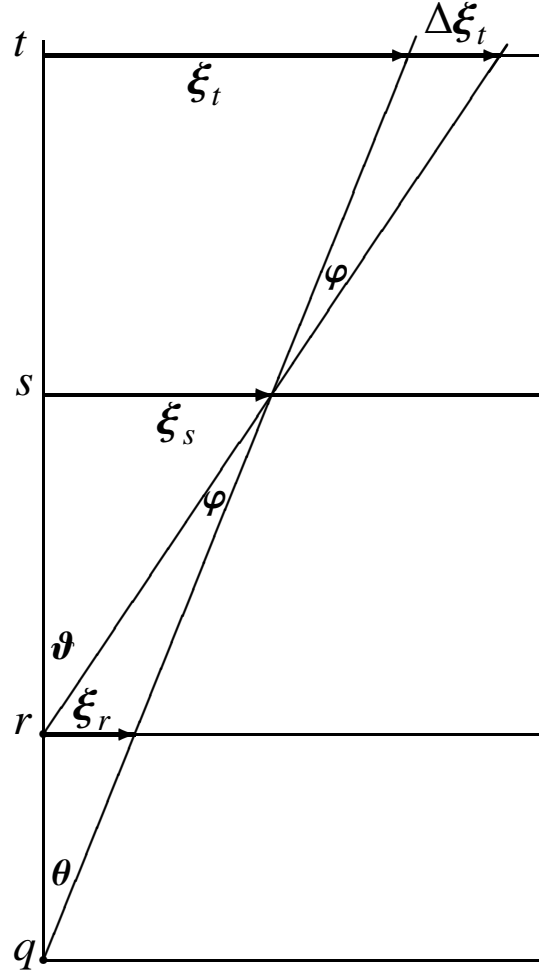
\begin{figure}
\setlength{\unitlength}{1.0cm}
\bc
\begin{picture}(7,13.4)
\thicklines
\put(0.5,0.5){\line(0,1){12.2}}
\put(0.5,0.5){\line(1,0){6.5}}
\put(0.5,3.5){\line(1,0){6.5}}
\put(0.5,8.0){\line(1,0){6.5}}
\put(0.5,12.5){\line(1,0){6.5}}
\put(0.5,3.5){\line(2,3){6.1}}
\put(0.5,0.5){\line(2,5){4.9}}

\put(0.5,3.51){\vector(1,0){1.2}}
\put(0.5,3.49){\vector(1,0){1.2}}
\put(0.5,8.01){\vector(1,0){3.0}}
\put(0.5,7.99){\vector(1,0){3.0}}
\put(0.5,12.51){\vector(1,0){4.8}}
\put(0.5,12.49){\vector(1,0){4.8}}
\put(5.3,12.49){\vector(1,0){1.2}}
\put(5.3,12.51){\vector(1,0){1.2}}

\put(0.1,0.5){\LARGE$q$}
\put(0.1,3.5){\LARGE$r$}
\put(0.1,8.0){\LARGE$s$}
\put(0.1,12.5){\LARGE$t$}
\put(0.9,3.7){\LARGE$\vc\xi_r$}
\put(2.0,7.5){\LARGE$\vc\xi_s$}
\put(2.4,12.0){\LARGE$\vc\xi_t$}
\put(5.5,12.8){\LARGE$\Delta\vc\xi_t$}
\put(4.2,9.5){\Large$\vc\vp$}
\put(2.6,6.5){\Large$\vc\vp$}
\put(0.6,1.5){\Large$\vc\theta$}
\put(0.6,4.5){\Large$\vc\vt$}

\put(0.5,0.5){\circle*{0.1}}
\put(0.5,3.5){\circle*{0.1}}
\end{picture}
\ec
\caption{Sketch of two light rays through four consecutive planes,
  with $0\le\lambda_q<\lambda_r<\lambda_s<\lambda_t$. The rays are not
  deflected in the lens planes. The first ray has its vertex at
  $\lambda_q$ and encloses an angle $\vc\theta$ with the fiducial ray;
  the second ray with vertex at $\lambda_r$ intersects the first ray
  at $\lambda_s$ and encloses an angle $\vc\vt$ with the fiducial
  ray. At the intersection point of the two rays, they enclose an angle
  $\vc\vp$. The geometry of this figure yields the relation
  (\ref{eq:Dgene}) between distance matrices}
\label{fig:1}
\end{figure}

Consider the pairs of light rays sketched in Fig.\ts\ref{fig:1}, where
the first has a vertex at $\lambda_q$ and encloses an angle
$\vc\theta$ with the fiducial ray. At the affine parameter
$\lambda_s$, its separation vector from the fiducial ray is
$\vc\xi_s$. The second light ray has its vertex at $\lambda_r$ and
intersects the first ray at $\lambda_s$; this
then specifies its direction $\vc\vt$ relative to the fiducial ray. At
the intersection point, the two rays enclose an angle $\vc\vp$. From
the geometry of the figure, we find
\bea
\vc\theta\!\!\!&=&\!\!\!\D_{qr}^{-1}\vc\xi_r
=\D_{qs}^{-1}\vc\xi_s=\D_{qt}^{-1}\vc\xi_t\; 
;  \nonumber \\
\vc\vt\!\!\!&=&\!\!\!\D_{rt}^{-1}
\rund{\vc\xi_t+\Delta\vc\xi_t}=\D_{rs}^{-1}\vc\xi_s\;.
\elabel{thet}
\eea
We will use the latter equation to derive a relation between the
$\D$'s, by expressing all vectors in terms of $\vc\xi_s$. Using the
first of (\ref{eq:thet}), we get $\vc\xi_t=\D_{qt}\D_{qs}^{-1}\xi_s$.
Furthermore, the figure shows that $\Delta\vc\xi_t=\D_{st}\vc\vp$. On
the other hand, $\vc\xi_r=-\D_s(\lambda_r)\vc\vp$.

We now have to relate $\D_s(\lambda_r)$, the backward extension of the
solution $\D_s(\lambda)$ of (\ref{eq:I6}), to $\D_{rs}$. For
that, we consider two solutions of (\ref{eq:I6}), $\D_r(\lambda)$
and $\D_s(\lambda)$, with their appropriate initial conditions at
$\lambda_r$ and $\lambda_s$, respectively, and define the matrix 
\be
\W(\lambda)={\d \D_r^{\rm t}\over \d\lambda}(\lambda)\; \D_s(\lambda)
-\D_r^{\rm t}(\lambda)\; {\d\D_s \over \d\lambda}(\lambda)\;,
\elabel{Wronsk}
\ee
which is the Wronskian of (\ref{eq:I6}); here, the superscript
`t' denotes the transpose of a matrix. The derivative of $\W$
vanishes, due to (\ref{eq:I6}); hence, $\W$ is a constant. 
Evaluating (\ref{eq:Wronsk}) at $\lambda=\lambda_r$ and making use of
the initial conditions of $\D_r(\lambda)$ yields
$\W=(1+z_r) \,\D_s(\lambda_r)$. Similarly, at $\lambda=\lambda_s$ we
find $\W=-(1+z_s)\, \rund{\D_r^{\rm
    t}}(\lambda_s)=-(1+z_s)\,\D_{rs}^{\rm t}$. Thus, we obtain
\be
\D_s(\lambda_r)=-{1+z_s \over 1+z_r}\,\D_{rs}^{\rm t}\;,
\elabel{Ether}
\ee
which is Etherington's theorem in matrix form \citep{Ethering33}.
With this relation, we then find that
\[
\Delta\vc\xi_t=\D_{st}\vc\vp=-\D_{st}\D_s^{-1}(\lambda_r)\,\vc\xi_r
={1+z_r \over 1+z_s}\,\D_{st} \rund{\D_{rs}^{\rm t}}^{-1}\vc\xi_r\;.
\]
Using (\ref{eq:thet}) and collecting terms,
\[
\D_{rt}\D_{rs}^{-1}\vc\xi_s=
\vc\xi_t+\Delta\vc\xi_t=\D_{qt}\D_{qs}^{-1}\vc\xi_s
+{1+z_r \over 1+z_s}\,\D_{st} \rund{\D_{rs}^{\rm
    t}}^{-1}\D_{qr}\D_{qs}^{-1} \vc\xi_s
\]
follows.
Since this relation is valid for
all $\vc\xi_s$, a general relation between distance matrices is
obtained: 
\be
{1+z_r \over 1+z_s}\,\D_{st} \rund{\D_{rs}^{\rm
    t}}^{-1}\D_{qr}=\D_{rt}\D_{rs}^{-1}\D_{qs}-\D_{qt}\;,
\elabel{Dgene}
\ee
where we multiplied the resulting equation by $\D_{qs}$ from the
right.\footnote{We explicitly point out that the only geometrical
  relation used in this derivation is the one between angles and
  transverse separations, i.e., the definition of the distance
  matrices.} 

Indeed, a relation of this kind is expected to hold: Consider
$\lambda_t\equiv \lambda$ as a variable. The two matrix-valued
functions $\D_q(\lambda)$ and $\D_r(\lambda)$ are linearly independent
solutions of the transport equation (\ref{eq:I6}), provided
$\lambda_r\ne \lambda_q$. Therefore, the solution $\D_s(\lambda)$ can
be written as a linear combination of the other two. This combination
should be of the form 
\be
\D_s(\lambda)=\eck{\D_r(\lambda)\D_r^{-1}(\lambda_s)-
\D_q(\lambda)\D_q^{-1}(\lambda_s)} \tens{X}\;,
\ee
which satisfies one of the initial conditions,
$\D_s(\lambda_s)=0$. The matrix $\tens{X}$ is determined from the
second initial condition; our result (\ref{eq:Dgene}) shows that 
\be
\D_s(\lambda)=\eck{\D_r(\lambda)\D_r^{-1}(\lambda_s)-
\D_q(\lambda)\D_q^{-1}(\lambda_s)} {1+z_s\over 1+z_r}
\D_{qs} \D_{qr}^{-1}\D_{rs}^{\rm t} \,.
\ee

\subsection{Equivalence of (\ref{eq:LEorig}) and (\ref{eq:LEnew1})} 
We shall now show that the two forms (\ref{eq:LEorig}) and
(\ref{eq:LEnew1}) of the lens equation are equivalent. As a first
step, we rewrite (\ref{eq:LEnew1}) in a form that admits a simple
geometrical interpretation. Specializing (\ref{eq:Dgene}) to $q=0$,
$r=j-1$, $s=j$, $t=j+1$ yields
\be
{1+z_{j-1} \over 1+z_j}\,\D_{j,j+1} \rund{\D_{j-1,j}^{\rm
    t}}^{-1}\D_{j-1}=\D_{j-1,j+1}\D_{j-1,j}^{-1}\D_{j}-\D_{j+1}\;.
\elabel{Dgene2}
\ee
We next consider the prefactor of $\vc\xi_j$ in
(\ref{eq:LEnew1}). Using
$\C_{j-1,j}^{-1}\D_{j-1,j}^{-1}=\D_{j-1}\D_j^{-1}$, which is obtained
from the definition (\ref{eq:Cdef}) of $\C$, we find that this
prefactor becomes
\be
\D_{j+1}\D_j^{-1}+{1+z_{j-1}\over 1+z_j}\D_{j,j+1}
\rund{\D_{j-1,j}^{-1}}^{\rm t}\D_{j-1}\D_j^{-1}
=\D_{j-1,j+1}\D_{j-1,j}^{-1}\;,
\elabel{DD}
\ee
where in the last step we made use of (\ref{eq:Dgene2}), and the fact
that the inversion and transposition operations on a matrix
commute. Thus, we can 
rewrite (\ref{eq:LEnew1}) in the form
\bea
\vc\xi_{j+1}\!\!\!&=&\!\!\!\D_{j-1,j+1}\D_{j-1,j}^{-1} \vc\xi_j
\nonumber \\
\!\!\!&-&\!\!\!
{1+z_{j-1}\over 1+z_j}\D_{j,j+1}\rund{\D_{j-1,j}^{-1}}^{\rm
  t}\vc\xi_{j-1}-\D_{j,j+1}\hat{\vc\alpha}_j(\vc\xi_j) \;.
\elabel{LEnew2}
\eea
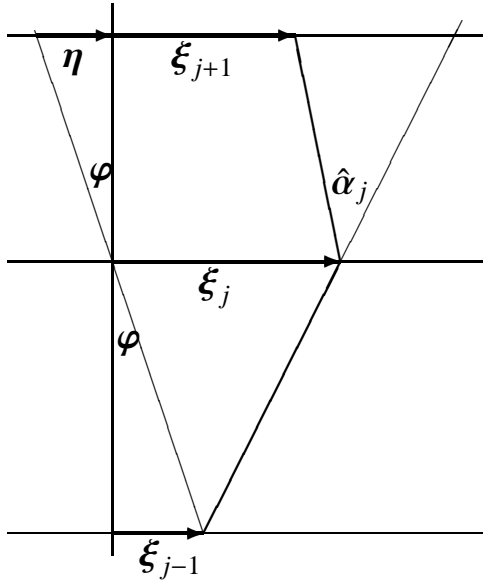
\begin{figure}
\setlength{\unitlength}{0.6cm}
\bc
\begin{picture}(11.2,12.5)
\thinlines
\put(0.7,1){\line(1,0){10.4}}
\put(0.7,7){\line(1,0){10.4}}
\put(0.7,12){\line(1,0){10.4}}
\put(3.0,0.5){\line(0,1){12.2}}
\put(5.0,1){\line(1,2){5.667}}
\put(5.0,1){\line(-1,3){3.7}}
\thicklines
\put(5.0,1){\line(1,2){3.0}}
\put(8.0,7){\line(-1,5){1.}}

\put(3.0,1.01){\vector(1,0){2.0}}
\put(3.0,0.99){\vector(1,0){2.0}}
\put(3.0,7.01){\vector(1,0){5.0}}
\put(3.0,6.99){\vector(1,0){5.0}}
\put(3.0,12.01){\vector(1,0){4.0}}
\put(3.0,11.99){\vector(1,0){4.0}}
\put(1.33,12.01){\vector(1,0){1.667}}
\put(1.33,11.99){\vector(1,0){1.667}}

\put(3.6,0.3){\Large$\vc\xi_{j-1}$}
\put(4.9,6.3){\Large$\vc\xi_j$}
\put(4.3,11.3){\Large$\vc\xi_{j+1}$}
\put(1.9,11.3){\Large$\vc\eta$}
\put(7.8,8.5){\Large$\hat{\vc\alpha}_j$}
\put(2.5,8.8){\Large$\vc\vp$}
\put(3.1,5.1){\Large$\vc\vp$}
\end{picture}
\ec
\caption{Propagation of a light ray (thick bent line) between three
  consecutive 
  planes. The vertical line is the optical axis, with respect to which
  the separation vectors $\vc\xi$ are measured. The geometry of this
  figures yields the lens equation (\ref{eq:LEnew2}) -- see text}
\label{fig:2}
\end{figure}

We note that this generalizes Eq.\ts(4.47) of SSE to the case of
general distance matrices between main lens planes.
This form of the lens equation can be immediately interpreted
geometrically. For this, we consider Fig.\ts\ref{fig:2}, from which we
read off
\[
\vc\eta+\vc\xi_{j+1}=\D_{j-1,j+1}
\D_{j-1,j}^{-1}\vc\xi_j-\D_{j,j+1}\hat{\vc\alpha}_j \;.
\]
Furthermore, $\vc\eta=\D_{j,j+1}\vc\vp$; on the other hand, 
$\vc\xi_{j-1}=-\D_{j,j-1}\vc\vp$. Eliminating $\vc\vp$ from these two
relations and making use of (\ref{eq:Ether}), we find 
\[
\vc\eta={1+z_{j-1} \over 1+z_j}\,\D_{j,j+1}\rund{\D_{j-1,j}^{-1}}^{\rm
  t}\vc\xi_{j-1} \;.
\]
Together, these two equations reproduce (\ref{eq:LEnew2}). In this
form, the equation not only is confined to three consecutive lens
planes, but all distance matrices occurring here are those between
these three planes. 

In order to show the equivalence of (\ref{eq:LEorig}) and
(\ref{eq:LEnew1}), it is useful to rewrite the prefactor of
$\vc\xi_{j-1}$ in (\ref{eq:LEnew2}) in a different form. Making use
again of (\ref{eq:Dgene2}), we obtain
\bea
\vc\xi_{j+1}\!\!\!\!\!&=&\!\!\!\!\!\D_{j-1,j+1}\D_{j-1,j}^{-1} \vc\xi_j
\elabel{LEnew3}
 \\
\!\!\!&+&\!\!\!\!\!
\rund{\D_{j+1}D_{j-1}^{-1}-\D_{j-1,j+1}\D_{j-1,j}^{-1}\D_j\D_{j-1}^{-1}}
\vc\xi_{j-1}-\D_{j,j+1}\hat{\vc\alpha}_j(\vc\xi_j) \;.
\nonumber 
\eea
We prove the equivalence by induction; for $j=1$, this equivalence is
seen by (\ref{eq:tau-1p}). Hence we assume that it is true for all
planes up to $j$. Then, taking the difference between
(\ref{eq:LEorig}) for $\xi_{j+1}$ and (\ref{eq:LEnew3}),
\bea
\Delta\!\!\!\!\!&=&\!\!\!\!\!\D_{j-1,j+1}\D_{j-1,j}^{-1} \vc\xi_j
\nonumber
 \\
\!\!\!\!\!&+&\!\!\!\!\!
\rund{\D_{j+1}D_{j-1}^{-1}-\D_{j-1,j+1}\D_{j-1,j}^{-1}\D_j\D_{j-1}^{-1}}
\vc\xi_{j-1}-\D_{j,j+1}\hat{\vc\alpha}_j(\vc\xi_j)\nonumber \\
\!\!\!\!\!&-&\!\!\!\!\!
\D_{j+1}\, \vc\theta+\sum_{i=1}^{j}
\D_{i,j+1}\,\hat{\vc\alpha}_i(\vc\xi_i) \;,
\eea
we need to show that $\Delta=0$. We first replace $\vc\xi_{j-1}$ and
$\vc\xi_j$ by their expressions from (\ref{eq:LEorig}), which holds
because of the induction assumption,
\bea
\Delta\!\!\!\!\!&=&\!\!\!\!\!\D_{j-1,j+1}\D_{j-1,j}^{-1} \rund{\D_j\vc\theta
-\sum_{i=1}^{j-1} \D_{i,j}\,\hat{\vc\alpha}_i(\vc\xi_i)}
\nonumber \\
\!\!\!\!\!&+&\!\!\!\!\!
\rund{\D_{j+1}-\D_{j-1,j+1}\D_{j-1,j}^{-1}\D_j}
\rund{\vc\theta- \sum_{i=1}^{j-2} \D_{j-1}^{-1}\D_{i,j-1}
\,\hat{\vc\alpha}_i(\vc\xi_i)} \\
\!\!\!\!\!&-&\!\!\!\!\!\D_{j,j+1}\hat{\vc\alpha}_j(\vc\xi_j)-
\D_{j+1}\, \vc\theta+\sum_{i=1}^{j}
\D_{i,j+1}\,\hat{\vc\alpha}_i(\vc\xi_i) \;.\nonumber
\eea
From this equation, one sees immediately that the terms
$\propto\vc\theta$ cancel each other. Second, the two terms $\propto
\hat{\vc\alpha}_j$ add up to zero. Third, also the sum of the two
terms $\propto \hat{\vc\alpha}_{j-1}$ is zero. Thus, what remains to be
shown is that the prefactor of the terms $\propto\hat{\vc\alpha}_i$,
\bea
K_i\!\!\!&=&\!\!\!
\D_{i,j+1}-\D_{j-1,j+1}\D_{j-1,j}^{-1}\D_{i,j} \nonumber \\
\!\!\!&-&\!\!\!
\rund{\D_{j+1}-\D_{j-1,j+1}\D_{j-1,j}^{-1}\D_j}
\D_{j-1}^{-1}\D_{i,j-1} \;,
\eea
for $i\le j-2$ vanish. For this, we consider again (\ref{eq:Dgene}),
setting $r=j-1$, $s=j$, $t=j+1$, once with $q=0$, and once with
$q=i$. This then yields
\bea
{1+z_{j-1} \over 1+z_j}\,\D_{j,j+1}\!\!\!\!\!\!\!\!&&\!\!\!\!\!\! \rund{\D_{j-1,j}^{\rm
    t}}^{-1}  =
\rund{\D_{j-1,j+1}\D_{j-1,j}^{-1}\D_{j}-\D_{j+1}} \D_{j-1}^{-1}
\nonumber \\
\!\!\!&=&\!\!\!
\rund{\D_{j-1,j+1}\D_{j-1,j}^{-1}\D_{i,j}-\D_{i,j+1}} \D_{i,j-1}^{-1}
\;.
\elabel{Dgene3}
\eea
After multiplying by $\D_{i,j-1}$, we see that the final equality
shows that $K_i=0$, which proves that $\Delta=0$, and thus the
equivalence of the two forms of the lens equation. The equation
$K_i=0$ itself provides an interesting relation between distance
matrices. 

\section{Mass-sheet transformation}
In standard gravitational lensing, with a single deflector between the
source and observer, there is a transformation of the mass
distribution of the lens which keeps most observables invariant, the
mass-sheet transformation \citep[MST, see][]{FGS85}. Since this
transformation is accompanied by a uniform isotropic scaling in the
source plane, all magnifications are scaled by the same factor, so
that magnification (and thus observable flux) ratios are
unchanged. The MST changes the product of time delay and Hubble
constant, though, and the corresponding degeneracy can thus be broken
by measuring the time delay in lens systems, assuming the Hubble
constant to be known from other cosmological observations
\citep[see][and references therein]{SS13}.
S14 has recently shown that a MST also exists in the case of two lens
planes and two source planes. In this section, we will show that also
for perturbed gravitational lens systems, as considered in this paper,
such a MST does exist.

\subsection{Single main lens plane}
We start with the case of a single lens plane, using  the lens
equation (\ref{eq:LE-1pl}), and modify the deflection angle
$\hat{\vc\alpha}_1(\vc\xi_1)$ to the new form
\be
\hat{\vc\alpha}_1'(\vc\xi_1)=\lambda \hat{\vc\alpha}_1(\vc\xi_1)
+\tens{G}_1\,\vc\xi_1\;,
\elabel{MST1}
\ee
where $\lambda$ is a real number, and $\tens{G}_1$ is a
matrix.\footnote{The use of the same symbol $\lambda$ for the affine
  parameter and the MST parameter is due to the conventions in the
  literature, but should not lead to any confusion; in particular, in
  this section $\lambda$ is exclusively used as MST parameter.}
Throughout this section, a prime denotes a mass-sheet transformed
quantity.  Thus, the modified deflection angle is a scaled version of
the original one, plus a term linear in the impact vector. If
$\tens{G}_1$ is symmetric, this linear term corresponds to a tidal
matrix, i.e., adding a uniform mass sheet to the scaled lens mass
distribution, plus an external shear. The modified lens equation then
becomes
\be
\vc\xi_2'=\D_2\,\vc\theta-\D_{12}\, \hat{\vc\alpha}_1'(\vc\xi_1)
=\D_2\,\vc\theta-\D_{12}\eck{\lambda \hat{\vc\alpha}_1(\vc\xi_1)
+\tens{G}_1\,\D_1 \vc\theta}\;.
\elabel{MST2}
\ee
As for the orignal MST, we require that the modified impact vector
$\vc\xi_2'$ is related to the original one by a uniform, isotropic
scaling, $\vc\xi_2'=\nu_2 \vc\xi_2$, where $\nu_2$ is the scaling
factor. Thus we require
\be
\D_2\,\vc\theta-\D_{12}\eck{\lambda \hat{\vc\alpha}_1(\vc\xi_1)
+\tens{G}_1\,\D_1 \vc\theta}=\nu_2\eck{\D_2\,\vc\theta-\D_{12}\, 
\hat{\vc\alpha}_1(\vc\xi_1)} \;.
\elabel{BBtt}
\ee
In order to have the terms $\propto\hat{\vc\alpha}_1$ equal on both
sides of (\ref{eq:BBtt}), we need to set $\nu_2=\lambda$, as is also
the case for the 
MST in standard lensing -- the scaling of the source plane (here plane
number 2) is the same as that of the deflection angle. The remaining
terms are all $\propto\vc\theta$, and setting them equal on both side
leads to $\D_2-\D_{12}\tens{G}_1\D_1=\lambda \D_2$, or
\be
\tens{G}_1=(1-\lambda)\D_{12}^{-1}\D_2\D_1^{-1}=(1-\lambda)\C_{12}\;.
\elabel{MST3}
\ee
Since $C_{12}$ is symmetric \citep[see][]{Schn97}, $\tens{G}_1$ is
indeed a tidal matrix. This single-main plane MST was also derived by
\cite{McCully14}.  Thus, in a generalized gravitational lens
situation, the MST requires a shear in addition to a uniform mass
sheet.\footnote{In case the distance matrices are proportional to the
  unit matrix, the transformation reduces to the known one in standard
  lensing. Note that the transformation (\ref{eq:MST1}) implies the
  transformation $\hat\psi_1'(\vc\xi_1)=\lambda \hat\psi_1(\vc\xi_1)
  +\vc\xi_1^{\rm t}\tens{G}_1\vc\xi_1/2$ for the deflection
  potential. For isotropic distance matrices,
  $\tens{G}_1=(1-\lambda)D_2/(D_1 D_{12}){\cal I}$, so that
  $\hat\psi_1'(\vc\xi_1)=\lambda \hat\psi_1(\vc\xi_1) +(1-\lambda)(D_1
  D_2/D_{12})|\vc\theta|^2/2$. According to (\ref{eq:psipsi}), this
  then implies for the scaled deflection potential
  $\psi'(\vc\theta)=\lambda
  \psi(\vc\theta)+(1-\lambda)|\vc\theta|^2/2$, as in standard lens
  theory.}

\subsection{Two main lens planes}
We now consider a second lens plane at $\lambda_2$, with the source
plane being located at $\lambda_3$. The modified lens equation then
reads 
\be
\vc\xi_3'=\D_3\vc\theta-\D_{13}\hat{\vc\alpha}_1'(\vc\xi_1')
-\D_{23}\hat{\vc\alpha}_2'(\vc\xi_2') \;,
\elabel{MST4}
\ee
and we require the modified deflection angle $\hat{\vc\alpha}_2'$ to
be chosen such that the 3-plane is just uniformly scaled relative to
the original one, i.e., $\vc\xi_3'=\nu_3\vc\xi_3$. This condition then
yields 
\bea
\D_3\vc\theta\!\!\!&-&\!\!\!\D_{13}\eck{\lambda\hat{\vc\alpha}_1(\vc\xi_1)
+\tens{G}_1\vc\xi_1}-\D_{23}\hat{\vc\alpha}_2'(\vc\xi_2')\nonumber \\
\!\!\!&=&\!\!\!\nu_3\eck{\D_3\vc\theta-\D_{13}\hat{\vc\alpha}_1(\vc\xi_1)
-\D_{23}\hat{\vc\alpha}_2(\vc\xi_2)} \;.
\elabel{MST5}
\eea
In order to account for the term $\propto \hat{\vc\alpha}_2$ on the
r.h.s. of (\ref{eq:MST5}), the modified deflection has to be of the
form 
\bea
\hat{\vc\alpha}_2'(\vc\xi_2')
\!\!\!&=&\!\!\!
\nu_3\hat{\vc\alpha}_2(\vc\xi_2'/\lambda)
+\tens{G}_2\vc\xi_2' \nonumber \\
\!\!\!&=&\!\!\! \nu_3\hat{\vc\alpha}_2(\vc\xi_2'/\lambda)
+\lambda \tens{G}_2
\eck{\D_2\vc\theta-\D_{12}\hat{\vc\alpha}_1(\vc\xi_1)} \;.
\elabel{MST6}
\eea
This choice then yields equal terms $\propto \hat{\vc\alpha}_2$ on
both sides. Equating the terms $\propto \hat{\vc\alpha}_1$ leads to
the condition
$-\lambda\D_{13}+\lambda\D_{23}\tens{G}_2\D_{12}=-\nu_3\D_{13}$, or
\be
\tens{G}_2=(1-\nu_3/\lambda)\D_{23}^{-1}\D_{13}\D_{12}^{-1}\;.
\elabel{MST7}
\ee
Using the same arguments as in the Appendix of \cite{Schn97}, it is
straightforward to show that any combination of distance matrices of
the form 
\be
\D_{st}^{-1}\D_{rt}\D_{rs}^{-1}\quad\hbox{is symmetric,}
\elabel{MST8}
\ee
for $0\le\lambda_r<\lambda_s<\lambda_t$. Hence, $\tens{G}_2$ 
is symmetric, and thus corresponds to a tidal matrix. Equating the
terms $\propto \vc\theta$ in (\ref{eq:MST5}) then leads to
\be
(1-\nu_3)\D_3=(1-\lambda)\D_{13}\D_{12}^{-1}\D_2
+\lambda(1-\nu_3/\lambda)\D_{13}\D_{12}^{-1}\D_2\;,
\elabel{MST9}
\ee
which has the unique solution $\nu_3=1$. Thus, as is the case for the
standard multi-plane lens discussed in \cite{S14}, the MST does lead to no
scaling in the plane $j=3$. Therefore,
\be
\tens{G}_2=(1-1/\lambda)\D_{23}^{-1}\D_{13}\D_{12}^{-1}\;.
\elabel{MST10}
\ee
The implied scaling of the mass distribution in the plane $i=2$ that
follows from (\ref{eq:MST6}) is discussed in \cite{S14}; in short, the
surface mass density distribution giving rise to
$\hat{\vc\alpha}_2(\vc\theta_2)$ needs to be scaled in amplitude and
scale-length to yield a deflection
$\hat{\vc\alpha}_2(\lambda\vc\theta_2)$. 

\subsection{Arbitrary number of planes}
Here we will generalize the MST to an arbitrary number of source/lens
planes. It turns out that the lens equation in the form
(\ref{eq:LEnew3}) is better suited for that purpose. We write it in
the form 
\be
\vc\xi_{j+1}=\D_{j-1,j+1}\D_{j-1,j}^{-1}\vc\xi_j + \tens{B}_j\vc\xi_{j-1}
-D_{j,j+1}\hat{\vc\alpha}_j(\vc\xi_j)\;,
\elabel{MST11}
\ee
where $\tens{B}_j$ is the term in parenthesis in (\ref{eq:LEnew3}). We will
now assume a scaling $\vc\xi_j'=\nu_j \vc\xi_j$ in every plane, and 
set the scaled deflection angles to be
\be
\hat{\vc\alpha}_j'(\vc\xi_j')=\nu_{j+1}\hat{\vc\alpha}_j(\vc\xi_j'/\nu_j)
+\tens{G}_j\vc\xi_j'
=\nu_{j+1}\hat{\vc\alpha}_j(\vc\xi_j)+\nu_j\tens{G}_j\vc\xi_j\;.
\elabel{MST12}
\ee
Note that (\ref{eq:MST1}) and (\ref{eq:MST6}) are special cases of the
relation (\ref{eq:MST12}) for $j=1,2$, respectively, and $\nu_1=1$,
$\nu_2=\lambda$, $\nu_3=1$. Then, from $\vc\xi_j'=\nu_j \vc\xi_j$, we
obtain from (\ref{eq:MST11})
\bea
\vc\xi_{j+1}'\!\!\!&=&\!\!\!
\nu_j\D_{j-1,j+1}\D_{j-1,j}^{-1}\vc\xi_j + \nu_{j-1}\tens{B}_j\vc\xi_{j-1}
\nonumber \\
\!\!\!&&-\nu_{j+1}D_{j,j+1}
\hat{\vc\alpha}_j(\vc\xi_j)-\nu_jD_{j,j+1}\tens{G}_j\vc\xi_j
\elabel{MST13}\\
\!\!\!&=&\!\!\!
\nu_{j+1}\eck{\D_{j-1,j+1}\D_{j-1,j}^{-1}\vc\xi_j + \tens{B}_j\vc\xi_{j-1}
-D_{j,j+1}\hat{\vc\alpha}_j(\vc\xi_j)} \;.\nonumber
\eea
The terms $\propto \hat{\vc\alpha}_j$ cancel each other. The terms
$\propto \vc\xi_{j-1}$ yield the condition $\nu_{j+1}=\nu_{j-1}$, and
equating the terms $\propto \vc\xi_j$ yields
\be
\tens{G}_j=(1-\nu_{j+1}/\nu_j)D_{j,j+1}^{-1}\D_{j-1,j+1}\D_{j-1,j}^{-1}\;,
\elabel{MST13a}
\ee
which is symmetric according to (\ref{eq:MST8}) and thus represents a
tidal matrix.
Thus, we obtain $\nu_j=\lambda$ for $j$ even, and $\nu_j=1$ for $j$
odd. Correspondingly,
\bea
\tens{G}_j\!\!\!&=&\!\!\! (1-1/\lambda)D_{j,j+1}^{-1}\D_{j-1,j+1}\D_{j-1,j}^{-1}
\;\;{\rm for}\;j\;{\rm even}\;; \nonumber \\
\tens{G}_j\!\!\!&=&\!\!\! (1-\lambda)D_{j,j+1}^{-1}\D_{j-1,j+1}\D_{j-1,j}^{-1}
\;\;{\rm for}\;j\;{\rm odd}\;.
\elabel{MST14}
\eea
We note that (\ref{eq:MST3}) and (\ref{eq:MST7}) are special cases of
(\ref{eq:MST14}). 

Hence we find that the MST in multiple (lens and source) plane
gravitational lensing exhibits a curious behavior: The scaling factor in
every second plane is just unity, whereas it is $\lambda$ is the other
half of the planes. In particular that means that a `standard candle'
or `standard rod' in one of the planes with $j$ odd cannot be used to
break the degeneracy related to the MST, as the images of these
sources are unaffected by the MST.
The prefactor in the tidal matrices $\tens{G}_j$
are positive on every other plane, and negative on the remaining ones.
If one disregards the perturbations between lens planes, so that the
distance matrices $\D_{ij}$ reduce to angular diameter distances
$D_{ij}$, then the $\tens{G}_j$ become scalars proportional to the
density of uniform mass sheets; in this case, positive and negative
densities of these sheets alternate.

\subsection{Transformation of the time delay}
We will next consider how the MST affects the time delays. For that,
we assume to have a source on plane number $N+1$, with its light being
deflected in $N$ main lens planes. The corresponding 
LTTF is given in (\ref{eq:tau}), where $\vc\xi_{N+1}$ is the
position of the source in its source plane.

To obtain the corresponding function $c\tau'$ after the MST, we first
need to consider the transformation of the potential time delay. That
is, we need to find the transformed deflection potential
$\hat\psi_i'(\vc\xi_i')$, which needs to satisfy $\nabla_{\vc\xi_i'}
\hat\psi_i'(\vc\xi_i')=\hat{\vc\alpha}_i'(\vc\xi_i')$. For this, we
make the ansatz $\hat\psi_i'(\vc\xi_i')=a \hat\psi_i(\vc\xi_i'/\nu_i)
+\vc\xi_i^{\prime{\rm t}}\tens{G}_i\vc\xi_i'/2$. Taking the gradient
yields $\nabla_{\vc\xi_i'} \hat\psi_i'(\vc\xi_i')
=(a/\nu_i)\hat{\vc\alpha}_i(\vc\xi_i'/\nu_i)+\tens{G}_i\vc\xi_i'$. This
is seen to agree with $\hat{\vc\alpha}_i'(\vc\xi_i')$ in
(\ref{eq:MST12}), provided $a=\nu_i\nu_{i+1}=\lambda$. Thus,
\be
\hat\psi_i'(\vc\xi_i')=\lambda \hat\psi_i(\vc\xi_i'/\nu_i)
+{1\over 2}\vc\xi_i^{\prime{\rm t}}\tens{G}_i\vc\xi_i' \;.
\elabel{MST15}
\ee
We then obtain for the transformed LTTF
\bea
c \!\!\!\!\!\!&&\!\!\!\!\!\!\!\!\!\tau' =
\sum_{i=1}^N (1+z_i) 
 \Biggl[
{1\over 2} 
\rund{\nu_i\C_{i,i+1}\vc\xi_i - \nu_{i+1}\D_{i,i+1}^{-1}\vc\xi_{i+1}}^{\rm t}
\C_{i,i+1}^{-1}
\nonumber \\
&&\qquad \qquad \times \rund{\nu_i\C_{i,i+1}\vc\xi_i -
  \nu_{i+1}\D_{i,i+1}^{-1}\vc\xi_{i+1}}
\nonumber \\
\!\!\!&&-  \lambda\hat\psi_i(\vc\xi_i)- {\nu_i^2\over 2}
\vc\xi_i^{{\rm t}}\tens{G}_i\vc\xi_i 
\Biggr]  \;. 
\elabel{tauprime}
\eea
We will now show that 
\be
\tau'=
\lambda \tau +
F(\vc\xi_{N+1})\;,
\elabel{MST16}
\ee
which means that the transformed LTTF just scales by a
factor $\lambda$, {\it independent of the plane on which the source is
  located}, plus a function which only depends on the location of the
source, and thus cancels when considering time delays, i.e.,
differences between $\tau$ for pairs of multiple images of the
source. 

Taking the difference of $\tau'-\lambda\tau$, we first note that the
terms $\propto \hat\psi_i$ drop out. Second, we note that the terms
containing products of $\vc\xi_i$ and $\vc\xi_{i+1}$ also cancel, since
$\nu_i\nu_{i+1}=\lambda$. Thus we find that 
\bea
c(\tau'-\lambda\tau)\!\!\!&=&\!\!\!
\sum_{i=1}^N {(1+z_i)\over 2}\Biggl\{ \vc\xi_i^{\rm
  t}\eck{\rund{\nu_i^2-\lambda}\C_{i,i+1}-\nu_i^2\tens{G}_i}\vc\xi_i 
\nonumber \\
\!\!\!&+&\!\!\!
\rund{\nu_{i+1}^2-\lambda}\xi_{i+1}^{\rm t}
\rund{\D_{i,i+1}^{-1}}^{\rm t}\C_{i,i+1}^{-1}
\D_{i,i+1}^{-1}\vc\xi_{i+1} \Biggr\} \;.
\elabel{MST17}
\eea
The term $i=1$ of the first sum vanishes, since $\nu_1=1$, and
(\ref{eq:MST3}) holds. The final term ($i=N$)
of the second sum depends only on
the source position, and thus corresponds to the function
$F(\vc\xi_{N+1})$ previously mentioned. Since this term is of no
interest, we simply drop it from now on. Relabeling the index of the
second sum as $i\to i-1$, we then get 
\be
c(\tau'-\lambda\tau)=
\vc\xi_i^{\rm t}\eck{\sum_{i=2}^N {(1+z_i)\over 2}(\nu_i^2-\lambda)
\tens{K}_i}\vc\xi_i\;,
\elabel{MST18}
\ee
where the matrices $\tens{K}_i$ are given as 
\be
\tens{K}_i={1+z_{i-1}\over 1+z_i}
\rund{\D_{i-1,i}^{-1}}^{\rm t}\C_{i-1,i}^{-1}
\D_{i-1,i}^{-1}+\C_{i,i+1}
-\rund{\nu_i^2\over \nu_i^2-\lambda}\tens{G}_i \;.
\elabel{MST19}
\ee
Since, according to the definition (\ref{eq:Cdef}), $\C_{i-1,i}^{-1}
\D_{i-1,i}^{-1}=\D_{i-1}\D_i^{-1}$, we can rewrite $\tens{K}_i$ as 
\be
\tens{K}_i={1+z_{i-1}\over 1+z_i}
\rund{\D_{i-1,i}^{-1}}^{\rm t}\D_{i-1}\D_i^{-1}+\C_{i,i+1}
-\rund{\nu_i^2\over \nu_i^2-\lambda}\tens{G}_i \;.
\elabel{MST20}
\ee
From (\ref{eq:DD}), 
\be
{1+z_{i-1}\over 1+z_i}\rund{\D_{i-1,i}^{-1}}^{\rm t}\D_{i-1}\D_i^{-1}
=\D_{i,i+1}^{-1}\D_{i-1,i+1}\D_{i-1,i}^{-1}
-\D_{i,i+1}^{-1}\D_{i+1}\D_i^{-1} 
\ee
is obtained.
Noting that the final term is $\C_{i,i+1}$, we obtain with (\ref{eq:MST13a})
that 
\be
\tens{K}_i=\eck{1-\rund{\nu_i^2\over
    \nu_i^2-\lambda}\rund{1-{\nu_{i+1}\over \nu_i}}}
\D_{i,i+1}^{-1}\D_{i-1,i+1}\D_{i-1,i}^{-1}\;.
\ee
However, the prefactor vanishes: if $i$ is odd, $\nu_i=1$,
$\nu_{i+1}=\lambda$, and thus 
\[
1-\rund{\nu_i^2\over
    \nu_i^2-\lambda}\rund{1-{\nu_{i+1}\over \nu_i}}
=1-\rund{1\over 1-\lambda}(1-\lambda)=0\;.
\]
If $i$ is even, $\nu_i=\lambda$,
$\nu_{i+1}=1$, and
\[
1-\rund{\nu_i^2\over
    \nu_i^2-\lambda}\rund{1-{\nu_{i+1}\over \nu_i}}
=1-\rund{\lambda^2\over \lambda^2-\lambda}\rund{1-{1\over \lambda}}
=0\;.
\]
Therefore, $\tens{K}_i=0$ for all $i$, which completes our
proof of the validity of (\ref{eq:MST16}). Since the result is
independent of the plane on which the 
source is located -- the time delay scales for all source planes with
$\lambda$ -- we see that all time delays are scaled by the factor
$\lambda$ under the MST. In particular this implies that the
degeneracy due to the MST cannot be broken from measuring time delay
ratios. 

\section{Discussion}
In this paper we have considered several aspects of the generalized
multi-plane gravitational lensing equation. In contrast to the
treatment in \cite{SEF} and more recent papers
\citep[e.g.,][]{McCully14}, we treat the light propagation between
main lens planes with a continuous formalism, offered by the optical
tidal equation, instead of slicing up the matter into several
`weak-lensing' lens planes.\footnote{Whereas these two treatments are
equivalent (indeed, as shown in SSE, the slicing into weak-lensing
planes corresponds to a discretized version of the optical tidal
equation), the continuous formalism is more convenient for
analytical calculations -- for example, obtaining a result such as 
(\ref{eq:Dgene}) using the discretized version is probably extremely
tedious.} For this, we made use of the
formalism of light propagation in arbitrary spacetimes, as given in
SSE. As a result, the distance matrices between lens planes are
not written in terms of recursion relations, but as solutions of the
optical tidal equation; the explicit solution in terms of an integral
over the tidal field caused by large-scale density inhomogeneities
along the line-of-sight is provided in (\ref{eq:I9}).

The time-delay function for generalized multi-plane lensing was
derived, using the same arguments as employed in \cite{SEF} for the
derivation of the time delay in ordinary multi-plane lensing. The
explicit form deviates from that obtained in \cite{McCully14}, in that
our result depends only on the impact vectors in the various main
planes, but not on the deflection angles. In other word, out
expression for $\tau$ yields the light travel time of a kinematically
possible ray with specified impact vectors $\vc\xi_i$ in the main lens
planes, up to an additive constant. Physical light rays are those for
which the light-travel time is stationary; this allows the derivation
of an iterative lens equation which relates the impact vectors of
three consecutive main lens planes to the deflection angle in the
middle one of those. We have shown that this form of the lens equation
is equivalent to the more standard one which contains the impact
vectors and deflection angles of all earlier lens planes. This
consecutive lens equation is probably preferable for the use in ray
tracing simulations \citep[see, e.g.][]{Petkova13}.\footnote{The
  advantage of (\ref{eq:LEnew1}) and (\ref{eq:LEorig}) is twofold:
  First, in order to calculate the impact vectors in all $N$ planes
  requires of order $N^2/2$ multiplications for each light ray when
  (\ref{eq:LEorig}) is used, compared to about $3N$ multiplications
  for (\ref{eq:LEnew1}). More significant, however, is the fact that
  (\ref{eq:LEnew1}) allows one to save on memory: whereas
  (\ref{eq:LEorig}) requires the information of all lens planes for
  each ray, one can treat with (\ref{eq:LEnew1}) a large set of rays,
  tracing them from plane to plane, and in each step require only the
  information on a single lens plane.}

Finally, we showed that the generalized multi-plane lensing admits a
mass-sheet transformation (MST) which leaves all observables but the time
delay invariant. In contrast to ordinary lensing, the MST corresponds
to adding a tidal matrix in each main lens plane. We obtained the
curious behavior that the uniform isotropic scaling of the source/lens
planes, which is the key aspect of the MST, alternates between planes;
in every second plane, the scaling corresponds to the MST parameter
$\lambda$, in the other half of the planes, the scaling is unity. In
particular, this implies that the magnification of sources living on
the odd planes with scaling factor unity, is unaffected by the MST. All
time delays -- i.e., for sources in all main planes -- scale as
$\lambda$ under the MST.

This curious behavior of the MST in multi-plane lensing may indeed
offer a way to break the corresponding degeneracy, at least in a
statistical way. As we discussed before, in the case of vanishing
perturbations between the main lens planes, the MST corresponds to
mass sheets of alternating sign from plane to plane. Since such a mass
sheet changes the slope of the total mass distribution, it means that
this slope change also alternates. If one now makes the perhaps
plausible assumption that the shape of the mean mass profiles of
lenses is the same, this alternating slope change would violate the
universality of the mean mass profile. Thus, in multi-plane lensing,
the mass-sheet generacy may be more easily lifted than in the case of
a single lens plane only.

We hope that the results obtained here will be useful for further
theoretical studies of generalized multi-plane lensing, as well as for
modeling lens systems in which more than one main deflector affects
the imaging properties between observer's sky and the source plane.

\begin{acknowledgements}
  The author thanks Thomas Collett, Dominique Sluse, and Sherry Suyu
  for helpful comments and discussions. This work was supported in
  part by the Deutsche Forschungsgemeinschaft under the TR33 `The Dark
  Universe'.
\end{acknowledgements}

\begin{appendix}
\section{Distance matrices in terms of peculiar gravitational
  potential} 
In this appendix we derive the expression (\ref{eq:I8}) for the
distance matrices in an inhomogeneous Universe. As usual in
cosmological weak lensing, we work in comoving coordinates, and
therefore replace the affine parameter $\lambda$ by the comoving
distance $\chi$, where $\d\chi/\d a=c/(a^2 H)$, and $a$ is the cosmic
scale factor normalized to unity today. From the Robertson--Walker
metric and the condition for null geodesics, we have $a\,\d\chi=-c\,\d
t$, and with (\ref{eq:I2}) follows $\d\chi=\d\lambda/a^2$. These
relations then imply that 
\bea
{\d\over \d\lambda}\!\!\!&=&\!\!\!{\d\chi\over \d\lambda}{\d\over \d\chi}
={1\over a^2}{\d\over \d\chi} \;, \nonumber \\
{\d^2\over \d\lambda^2}\!\!\!&=&\!\!\!
{1\over a^4}{\d^2\over \d\chi^2}-{2\over a^3}{H\over c}{\d\over \d\chi}
\elabel{A1}
\\
a{\d^2\over \d\lambda^2}\!\!\!&+&\!\!\!{2 H\over c}{\d\over \d\lambda}
={1\over a^3}{\d^2\over \d\chi^2} \nonumber 
\eea
The angular-diameter distance $D_i$ is related to the comoving angular
diameter distance by $D_i=a\,f_k(\chi-\chi_i)$, where $\chi_i$ is the
comoving distance corresponding to the affine parameter $\lambda_i$. 
Applying (\ref{eq:I4}), we get
\be
{\d^2 D_i\over\d\lambda^2}=a{\d^2 f_k\over \d\lambda^2}
+2{\d a\over \d\lambda}{\d f_k\over \d\lambda}
+{\d^2 a\over \d\lambda^2}f_k
=-{3\over 2}\rund{H_0\over c}^2{\Omega_{\rm m}\over a^4}f_k\;.
\elabel{A2}
\ee
With $\d a/\d\lambda=(\d\chi/\d\lambda)(\d a/\d\chi)=H/c$, and 
\bea
{\d^2 a\over \d\lambda^2}\!\!\!\!&=&\!\!\!\!{\d H\over c\,\d\lambda}
={\d a\over \d\lambda}{\d H\over c\,\d a}
={1\over 2 c^2}{\d H^2\over \d a} \nonumber \\
\!\!\!\!&=&\!\!\!\!{H_0^2\over 2 c^2}\rund{2{\Omega_{\rm
      m}+\Omega_\Lambda-1\over a^3} 
-{3 \Omega_{\rm m}\over a^4}}\;, \nonumber 
\eea
and making use of the relations (\ref{eq:A1}), (\ref{eq:A2}) reduces
to (\ref{eq:I9}).

Next we write the distance matrix as $\D_i=D_i\,\tens{B}_i$, so that
$\tens{B}_i$ describes the deviation of $\D_i$ from the unperturbed
distance matrix $D_i{\cal I}$.\footnote{We note that $\tens{B}\equiv
\tens{B}_0$ 
corresponds to the matrix $\vc B$ in \cite{McCully14}, and the
$\tens{B}_i$ corresponds, up to a prefactor, to their matrices $\vc
C$.} This factorization transforms
(\ref{eq:I6}) into
\be
{\d^2\tens{B}_i\over \d\lambda^2}D_i
+2{\d\tens{B}_i\over \d\lambda}{\d D_i\over \d\lambda}
+\tens{B}_i{\d^2 D_i\over \d\lambda^2}
=\rund{{\cal T}_{\rm bg}+{\cal T}_{\rm sm}}\tens{B}_i D_i\;.
\elabel{A3}
\ee
Subtracting from this the transport equation (\ref{eq:I4}) for $D_i$,
we are left with 
\be
{\d^2\tens{B}_i\over \d\lambda^2}D_i
+2{\d\tens{B}_i\over \d\lambda}{\d D_i\over \d\lambda}
={\cal T}_{\rm sm}\tens{B}_i D_i\;.
\elabel{A4}
\ee
Inserting $D_i(\lambda)=a(\lambda)\,f_k(\chi-\chi_i)$, and using the
differentiation rules (\ref{eq:A1}), this turns into
\be
{\d^2\tens{B}_i\over \d\chi^2}f_k(\chi-\chi_i)
+2{\d\tens{B}_i\over \d\chi}{\d f_k\over \d\chi}
={\cal T}_{\rm sm}\tens{B}_i a^4 f_k(\chi-\chi_i)\;.
\elabel{A5}
\ee
We this relation, we find for
$\tens{X}_i(\chi):=f_k(\chi-\chi_i)\tens{B}_i(\chi)$:
\be
{\d^2 \tens{X}_i\over \d\chi^2}
={\d^2\tens{B}_i\over \d\chi^2}f_k(\chi-\chi_i)
+2{\d\tens{B}_i\over \d\chi}{\d f_k\over \d\chi}
-K\tens{X}=
-K \tens{X}_i +{\cal T}_{\rm sm} a^4 \tens{X}_i \;,
\elabel{A6}
\ee
where we made use of (\ref{eq:I9}). This differential equation can be
transformed into an integral equation, using the method of Green's
functions, to read
\be
\tens{X}_i(\chi)
=f_k(\chi-\chi_i){\cal I}
+\int_{\chi_i}^\chi\d\chi'\;f_k(\chi-\chi')
{\cal T}_{\rm sm}(\chi') a^4(\chi')
\tens{X}_i(\chi')\;.
\elabel{A7}
\ee
By differtiating twice, one can easily show that (\ref{eq:A7}) indeed
is a formal solution of (\ref{eq:A6}), with the correct initial
condition $\tens{X}_i(\chi_i)=0$, $\d \tens{X}_i(\chi_i)/\d\chi={\cal
  I}$. If we now use the expression (\ref{eq:I4a}) for ${\cal T}_{\rm
  sm}$, neglecting the final term which is a derivative along the
line-of-sight and thus cancels in the integration, and replacing the
derivatives w.r.t. $\vc\xi$ by those w.r.t. comoving transverse
coordinates, we get
\be
\tens{X}_i(\chi)
=f_k(\chi-\chi_i){\cal I}
-{2\over c^2}\int_{\chi_i}^\chi\d\chi'\;f_k(\chi-\chi')
\tens{H}(\phi(\chi'))
\tens{X}_i(\chi')\;.
\elabel{A8}
\ee
Using $\D_i= a\,\tens{X}_i$, we arrive at (\ref{eq:I8}). 
We also note that the large-scale structure component of the optical
tidal matrix can alternatively written in the form \citep[see][]{SS94}
\be
{\cal T}_{\rm sm}=-{2\over c^2\,a^4}
\eck{2\pi G a^2 (\rho-\bar\rho)+\rund{\begin{array}{cc}
\Gamma_1 & \Gamma_2 \\
\Gamma_2 & -\Gamma_1
\end{array} }} \;,
\elabel{A9}
\ee
with $\Gamma_1=(\phi_{,11}-\phi_{,22})/2$ and $\Gamma_2=\phi_{,12}$,
where the partial derivatives are with respect to transverse comoving
coordinates. Thus, one can replace $\tens{H}(\phi)$ in (\ref{eq:A8})
by the bracket in (\ref{eq:A9}).

\end{appendix}

\bibliographystyle{aa}
%\bibliography{/Users/sluse/work/articles/bibds}
\bibliography{MSDbib}

%\begin{thebibliography}{}

%\end{thebibliography}

\end{document}